\documentclass[prl,longbibliography,floatfix, twocolumn]{revtex4-1}
\usepackage{latexsym}
\usepackage{lineno}
\usepackage{hyperref}
\usepackage{url}
\usepackage[caption=false]{subfig}
\usepackage{graphicx}
\usepackage{verbatim}
\usepackage{multirow}
\usepackage{amsmath, bm}
\newcommand{\beq}{\begin{eqnarray}}
\newcommand{\eeq}{\end{eqnarray}}
\usepackage{mathrsfs}
\usepackage{float}
\usepackage{dsfont}
\usepackage{bbm}
\usepackage[usenames, dvipsnames]{color}
\usepackage{mathtools}
\usepackage{slashed}
\usepackage{physics}	% installed packages for typing maths and physics
\usepackage{graphicx}   % need for Fig.ures
\usepackage{epstopdf}
\usepackage{hyperref}   % use for hypertext links, including those to external documents and URLs
\usepackage{bbold}
\usepackage{wasysym}
\usepackage{amssymb}
\usepackage{feynmp}
\usepackage[symbol]{footmisc}
\usepackage[latin1]{inputenc}
\usepackage{tikz}
\usetikzlibrary{decorations.pathmorphing}
\usetikzlibrary{shapes.misc}
\tikzset{cross/.style={cross out, draw=black, minimum size=8*(#1-\pgflinewidth), inner sep=0pt, outer sep=0pt},
cross/.default={1pt}}
\usetikzlibrary{patterns,math}
\newcommand{\RN}[1]{%
\textup{\uppercase\expandafter{\romannumeral#1}}%
}

\newcommand{\ncmd}{\newcommand}
\ncmd{\nn}{\nonumber}
\ncmd{\mbf}[1]{\bs{#1}}
\ncmd{\gam}{\gamma}
\ncmd{\sig}{\sigma}
\ncmd{\pha}{\alpha}
\ncmd{\lam}{\lambda}
\ncmd{\dl}{\delta}
\ncmd{\kap}{\kappa}
\ncmd{\Lam}{\Lambda}
\ncmd{\Gam}{\Gamma}
\ncmd{\Dl}{\Delta}
\ncmd{\Ups}{\Upsilon}
\ncmd{\Om}{\Omega}
\ncmd{\eps}{\epsilon}
\ncmd{\veps}{\varepsilon}
\ncmd{\vphi}{\varphi}
\ncmd{\vtheta}{\vartheta}
\ncmd{\tw}{\text{w}}
\ncmd{\pll}{\parallel}
\ncmd{\mc}{\mathcal}
\ncmd{\mf}{\mathfrak}
\ncmd{\bs}{\boldsymbol}
\usepackage{pifont}
\usepackage[T1]{fontenc}
\usepackage{bclogo}
\ncmd{\note}[1]{{\color{red}{\ding{168} #1}}}
\ncmd{\eq}[1]{Eq. \eqref{#1}}
\ncmd{\fig}[1]{Fig. \ref{#1}}
\ncmd{\suppl}{\note{`Supplementary Information'}}
\ncmd{\pg}[1]{\textcolor{red}{#1}}

\begin{document}
\title{
Symmetry constraints and spectral crossing in a  Mott insulator \\
with Green's function zeros
}
\author{Chandan Setty$^{\oplus,*,1}$, Shouvik Sur$^{\dagger,*,1}$, Lei\ Chen$^{1}$, Fang\ Xie$^{1}$, Haoyu\ Hu$^2$, 
Silke\  Paschen$^{3,1}$, 
Jennifer Cano$^{4,5}$, and Qimiao Si$^{1}$}
\affiliation{$^1$Department of Physics and Astronomy, Rice Center for Quantum Materials, Rice University, Houston, Texas 77005, USA}
\affiliation{$^2$Donostia International Physics Center, P. Manuel de Lardizabal 4, 20018 Donostia-San Sebastian, Spain}
\affiliation{$^3$Institute of Solid State Physics, Vienna University of Technology, Wiedner Hauptstr. 8-10, 1040, Vienna, Austria}
\affiliation{$^4$Department of Physics and Astronomy, Stony Brook University, Stony Brook, NY 11794, USA}
\affiliation{$^5$Center for Computational Quantum Physics, Flatiron Institute, New York, NY 10010, USA}
\begin{abstract}
Lattice symmetries are central to the characterization of electronic topology. Recently, it was shown that Green's function eigenvectors form a representation of the space group. This formulation has allowed the identification of gapless topological states  even when quasiparticles are absent. Here we demonstrate the profundity of the framework in the extreme case, when interactions lead to a Mott insulator, through a solvable model with long-range interactions. We find that both Mott poles and zeros are subject to the symmetry constraints, and relate the 
symmetry-enforced spectral crossings to degeneracies of the original non-interacting eigenstates. Our results 
lead to new understandings of topological quantum materials and
highlight the utility of interacting Green's functions toward 
their
symmetry-based design.
\end{abstract}

\maketitle
\paragraph*{{\bf Introduction: }} In 
band theory of non-interacting topological semimetals, lattice symmetries 
act as indicators of topology and have been widely exploited in
identifying
novel topological materials 
~\cite{Armitage2017,Nagaosa2020, Bradlyn2017,Cano2018,Po2017,Watanabe2017,cano2021band}.
 The effects of interactions in topological semimetals are typically analyzed perturbatively~\cite{Armitage2017, son2007,sun2009, vafek2010,herbut2009,sur2016, huh2016, roy2017, sur2019}.  
 Interacting systems have also been treated approximately 
  in terms of renormalized non-interacting Hamiltonians (the so called Topological Hamiltonians) \cite{Wang-Zhang_PRX2012,Wang-Yan2013},
  where lattice symmetries can constrain 
  single-particle \cite{Lessnich2021} and collective\cite{Soldini2022} excitations.
 To address the interplay between strong correlations and topology, however, non-perturbative
 approaches to the interactions are required.
Whether and how symmetry constraints operate 
is {\it a priori} unclear, 
{especially when the interaction terms in a Hamiltonian do not commute with the single-particle terms.}
\par

Recently, a group that includes several of us have shown that the Green's function eigenvectors form 
a representation of the space group~\cite{Hu-Si2021}.
%in parallel to the Bloch functions of the 
%non-interacting settings~\cite{cano2021band}.
Symmetry enforced or protected 
degeneracies then respectively follow when the dimensionality of irreducible representation is greater than one at 
a given high symmetry point, or when two irreducible representations with distinct symmetry eigenvalues cross along a high symmetry line.
This formulation was applied to the case of a multi-channel Kondo lattice,
which describes a non-Fermi liquid metal with dispersive modes arising from 
fractionalized electronic excitations.
The eigenvectors of the Green's function were used to define degeneracies by locating spectral crossings~\cite{Hu-Si2021}.
The approach also provided the theoretical basis for the robustness~\cite{Chen-Si2022} of Kondo-driven Weyl semimetals~\cite{Lai2018,Dzsaber2017,Dzs-giant21.1}.

The extreme form of correlation effects occurs when the interactions drive a metal into a Mott localized state. 
It is an intriguing question as to what role topological nodes of the non-interacting limit may have in Mott insulators~\cite{Nagaosa2016}.
Along this direction, determining how symmetry constraints operate in a Mott insulator represents an outstanding open question. One of the  important features of a
Mott insulator 
is that it can have Green's function poles and zeros, both of which contribute to the Luttinger count of electronic states~\cite{AGD}.
Does symmetry constrain both features?

%%%%%%%%%%%%%%%%%%%%%%%%%%%%%%%
 \begin{figure*}[!t]
    \centering
\includegraphics[width=0.9\linewidth]{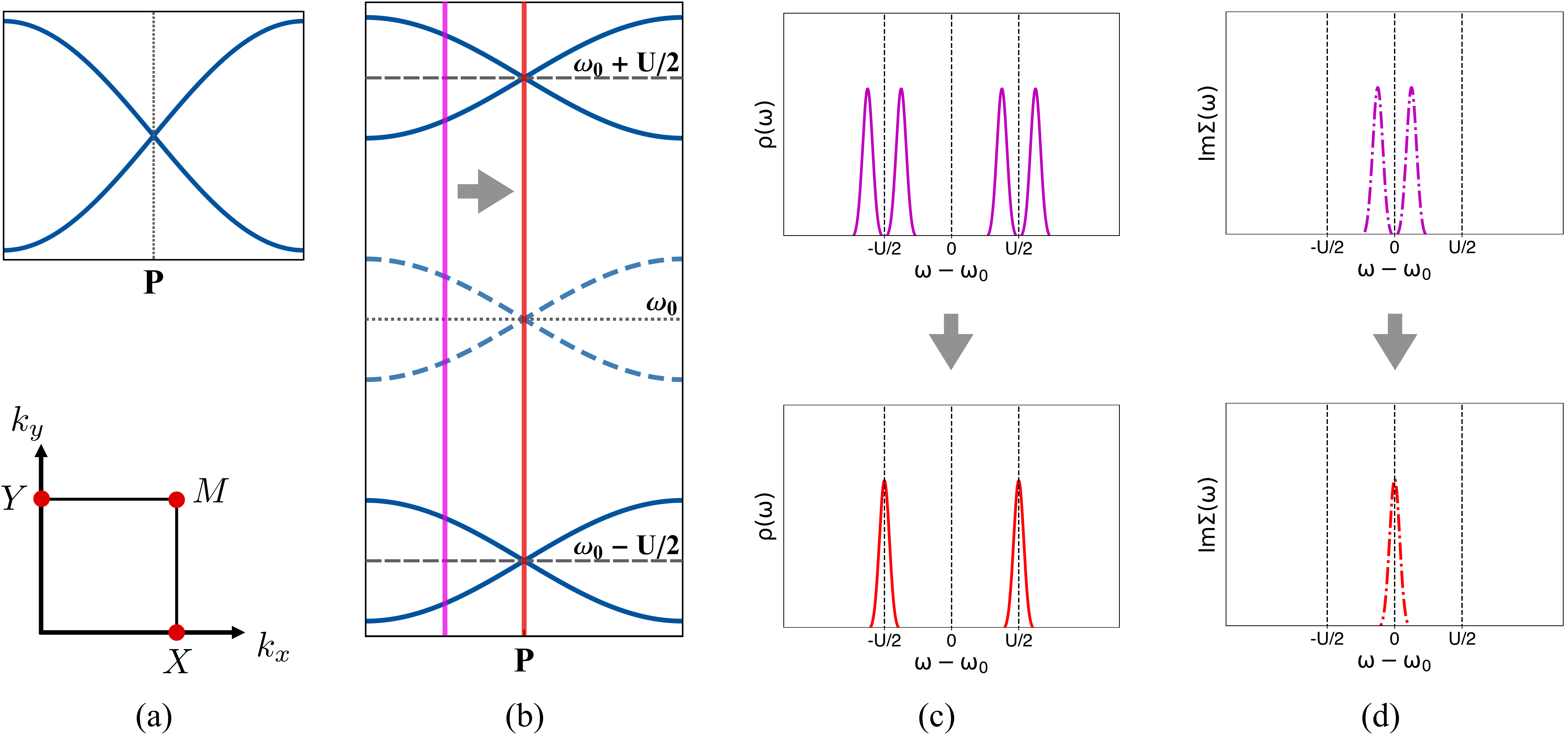}
    \caption{ Schematic summary of symmetry constraints and spectral crossings in a Mott insulator with Green's functions zeros and poles (a) Top: Symmetry enforced  Dirac point
in the non-interacting dispersion that occurs at a high symmetry point $P$. Bottom: Dirac points (red dots) that occur in the square net lattice at high symmetry points in the Brillouin zone. 
    (b) Symmetry enforced spectral crossings
    in the Mott insulator state. 
    They involve the upper
    and lower Hubbard bands (solid curves) and their associated Dirac points are separated by $U$ and a contour of crossings of zeros (dashed curves). Both are enforced by
    the lattice symmetry. 
    (c) Top: Spectral function (imaginary part of the Green's function) at a wavevector marked by the magenta line in (b) away from the Dirac point for the poles of the Green's function. Bottom: Spectral function
   at the wavevector $P$ marked by
    the red line in (b) at the Dirac point.  
    (d) Top: Imaginary part of the self energy 
    at a wavevector marked by the magenta line in (b) indicating zeros of the Green's function. Bottom: Same 
    at the wavevector $P$ marked by the red line indicating degeneracies of zeros enforced by symmetry. 
    }
    \label{Fig:Schematic}
\end{figure*}
%%%%%%%%%%%%%%%%%%%%%%%%%%%%%%%

In this work, 
we 
address the symmetry constraints of a Mott insulator using the Green's function approach~\cite{Hu-Si2021}. To be specific, we 
present our analysis
on 
a lattice model in which the non-interacting Hamiltonian has symmetry-enforced Dirac nodes, 
though we expect our results to be valid more generally.
Importantly, the symmetry constrains the Green's functions at all frequencies and the degeneracies
at the high symmetry wavevectors appear in the form of spectral crossings; in particular, we find that this operates
on both Green's function poles and zeros.
Our qualitative results are illustrated in Fig.\,\ref{Fig:Schematic}:
the spectral crossings of the Green's function poles [(c)] and Green's function zeros [(d)] appears as the wavector moves [(b)] towards the high symmetry wavevector $P$;
this captures the degereracy of the Green's function eigenvectors at $P$, where the Bloch functions of the  non-interacting counterpart are degenerate [(a), top panel].
 They
give rise to new understandings of topological quantum materials and
 set
the stage for systematic analysis of the topology of Mott insulators.
\par

 {\bf Interacting square net lattice and solution method:}
 We consider a two-dimensional (2D) square net lattice, 
 as illustrated in  Fig.~1 in the Supplementary Material (SM)~\cite{sm}.
Here, the
  non-interacting bands contain symmetry enforced Dirac crossings at the $X$ and $M$ points in the Brillouin zone 
  (Fig.\,\ref{Fig:Schematic}(a), bottom panel)
  ~\cite{Young-Kane2015}.
  We focus on local in momentum interactions analogous to those
  appearing in the Hatsugai-Kohmoto (HK) model~\cite{HK1992,Nagaosa2016, Setty2018, Setty2020, Setty2021, Phillips2020, Phillips2021, Yang2021, Wang2021, Setty2021-Kondo,Fabrizio2022}. 
  This form of interaction can be solved exactly [see the SM~\cite{sm}, Sec.~III], which facilitates the understanding of not only the symmetry-enforced spectral crossing but also the symmetry constraints on dispersive poles and zeros as we do below.

The Hamiltonian of a 2D 
square net lattice
(see SM~\cite{sm})
in the 
orbital basis
$\Lambda_{\bs k}^\intercal \equiv (
c_{A, \Uparrow}, c_{A, \Downarrow}, c_{B, \Uparrow}, c_{B, \Downarrow}
)_{\bs k} 
$
takes the form $\tilde{\mathscr H} = \tilde{\mathscr H}_0 + \tilde{\mathscr H}_I$, where 
\begin{align}
&\tilde{\mathscr H}_0 =  \sum_{\bs k} \Lambda_{\bs k}^{\dagger} \tilde h_0(\bs k) \Lambda_{\bs k} \nn \\
& \tilde{\mathscr H}_I =  \frac{\alpha}{2} \sum_{\bs k}  \Lambda_{\bs k}^\dagger  \Lambda_{\bs k} + \frac{U_c}{2} \sum_{\bs k} \left( \Lambda_{\bs k}^\dagger  \Lambda_{\bs k}\right)^2  \nonumber \\
& \qquad  + \frac{U_s}{2} \sum_{\bs k} \qty( \Lambda_{\bs k}^\dagger \tau_1 \otimes \sigma_0 \Lambda_{\bs k})^2
\end{align}
with  
$\tilde{h}_0 (\bs k) = t_2 (\cos{k_x} + \cos{k_y})  \mathbbm{1} + 2t \cos{\frac{k_x}{2}} \cos{\frac{k_y}{2}} \tau_1 \otimes \sigma_0 + t_{SO} \left[ \sin{k_x} \tau_3 \otimes \sigma_2 - \sin{k_y} \tau_3 \otimes \sigma_1 \right] - \mu \mathbbm{1}$, $\tau_j$  ($\sigma_j$) being the $j$-th Pauli matrix acting on the sublattice (spin) subspace, and $\mathbbm{1}$ is the $4\times 4$ identity matrix.   
Here, $c_{a, b}$ are the annihilation operators for electrons in the original sublattice $a = A, B$ and physical spin $b = \Uparrow, \Downarrow$ indices. 
$t, t_2$ are the hopping parameters, $t_{SO}$ is the spin-orbit coupling, $U_c$ and $U_s$ are intra- and inter-orbital interaction, respectively, 
and $\alpha$ is a constant shift in the density which we fix to $-(U_c+ U_s)$ for convenience.
{As we shall show below, a finite $U_s$ introduces an asymmetry between the dispersion of the poles and zeros of the interacting Greens function, which is expected in a generic interacting system.}
Without loss of generality, henceforth, we set $t_2=0$; {the non-interacting bandwidth is $8t$}. %%

It is convenient to work in a basis where the non-interacting Hamiltonian is block diagonal. To this end, we rotate the original basis of $\Lambda_{\bs k}^\intercal$ into the new basis $\Phi_{\bs k}^\intercal \equiv (
\phi_{+, \Uparrow}, \phi_{+, \Downarrow}, \phi_{-, \Uparrow}, \phi_{-, \Downarrow}
)_{\bs k} \equiv \frac{1}{\sqrt{2}}
( c_{A, \Uparrow} + c_{B, \Uparrow},
c_{A, \Downarrow} - c_{B, \Downarrow},
- c_{A, \Uparrow} + c_{B, \Uparrow}, 
c_{A, \Downarrow} + c_{B, \Downarrow}
)_{\bs k} 
$. 
This amounts to block-diagonalizing the Hamiltonian matrix as  $\tilde{h}_0(\bs k)\to  h_0(\bs k) = e^{i \frac{\pi}{4} \tau_2 \otimes \sigma_3} \tilde{h}_0(\bs k) e^{-i \frac{\pi}{4} \tau_2 \otimes \sigma_3}$,
%\begin{align}
%\end{align}
%%
where $h_0(\bs k) =\vec n(\bs k) \cdot \vec \Gamma- \mu \mathbbm{1}$ with $\vec n(\bs k) = \qty{- t_{SO} \sin{k_y}, t_{SO} \sin{k_x}, 2t \cos{\frac{k_x}{2}} \cos{\frac{k_y}{2}} }$ and $\vec \Gamma = \tau_3 \otimes \vec \sigma$.
It supports doubly degenerate bands which disperse as $\tilde \xi_j = -\mu + (-1)^j |\vec n(\bs k)|$.
In  the $\Phi_{\bs k} $ basis, 
\beq
\label{Eq:InteractingHamiltonianOrbitalBasis}
\tilde{\mathscr H}_I \to {\mathscr H}_I  &=&  \frac{U_c}{2} \sum_{\bs k} \left( \Phi_{\bs k}^\dagger  \Phi_{\bs k}\right)^2 +\frac{\alpha}{2} \sum_{\bs k}  \Phi_{\bs k}^\dagger  \Phi_{\bs k}  \\ \nonumber 
&+& \frac{U_s}{2} \sum_{\bs k} \qty( \Phi_{\bs k}^\dagger \tau_3 \otimes \sigma_3 \Phi_{\bs k})^2. 
 \eeq
 Here $\Phi_{\bs k}^\dagger  \Phi_{\bs k}$
 is the total charge and 
 $ \Phi_{\bs k}^\dagger \tau_3 \otimes \sigma_3 \Phi_{\bs k}$
 is a staggered  pseudo-sublattice density for a given momentum.

In the band basis, $\Psi_{\bs k} = (
\psi_{1, \uparrow}, \psi_{1, \downarrow}, \psi_{2, \uparrow}, \psi_{2, \downarrow}
)_{\bs k}^\intercal$ with $\{\psi_{j, \uparrow}, \psi_{j, \downarrow}\}$ representing the $j$-th pair of degenerate bands, $h_0$ is diagonalized to  $\qty[- \mu \mathbbm{1} + |\vec n(\bs k)| \tau_3\otimes \sigma_0]$.
Henceforth, we treat $\sigma = \uparrow, \downarrow$ and $i, j =1,2$ as a pseudo-spin and band indices respectively.
In the limit of weak spin-orbit coupling, $(t_{SO}/t) \ll 1$, the total Hamiltonian can be cast into the form $H = H_0 + H_I$  with $H_0 = \sum_{\bs k i \sigma} \xi_{i \sigma}(\bs k) \psi^{\dagger}_{i\sigma}(\bs k)\psi_{i\sigma}(\bs k)$ and 
\beq
H_I = U \sum_{\bs k i} n_{\bs k i \uparrow} n_{\bs k i \downarrow}  + U' \sum_{\substack{\bs k \sigma \sigma'\\ i\neq j}} n_{\bs k i \sigma} n_{\bs k j \sigma'} + \order{t_{SO}/t},
\label{Eq:BandBasisHamiltonian}
\eeq
where $(U, U') = (U_c + U_s, U_c-U_s)$, and correspond to  intra-band and inter-band interactions respectively. 
We utilize the density basis to exactly diagonalize the interacting Hamiltonian with interaction terms up to $\order{(t_{SO}/t)^0}$. 
The additional $\order{(t_{SO}/t)^{n>0}}$ terms only  distort the bands keeping the degeneracies intact. 
A full numerical solution for $t \sim t_{SO}$ appears 
later in the main text and in the 
SM~\cite{sm} (Sec.~I).
The renormalized band dispersions satisfy $\xi_{i\uparrow}(\bs k) = 
\xi_{i \downarrow} (\bs k)$, and are related to the bare band dispersions, $\tilde{\xi}_i (\bs k
)$, as $\xi_i(\bs k) \equiv \tilde{\xi}_i(\bs k) - U/2$. The density operators of $\psi_{\bs k i \sigma}$ are denoted as $ n_{\bs k i \sigma}$. 
\par
In the presence of time reversal symmetry, the total Green's function can be evaluated exactly as outlined in  
the SM~\cite{sm} (Sec.~II).
The calculation captures Fig.~3 in SM~\cite{sm}, which shows the transitions that contribute to the zero of the Green's function.
It simplifies in the zero temperature limit ($\beta \rightarrow \infty$) where both $\xi_1(\bs k), \xi_2(\bs k)$ are filled with $U, U'>0$. 
Further, for each $\bs k$, when $U + 2 U'> |\xi_1| + 2 |\xi_2|, 2|\xi_1| + |\xi_2|$ and $U>2|\xi_1|, 2 |\xi_2|$ but $U'< |\xi_1| + |\xi_2|$, the partition function is $Z_{\bs k} = \lim_{\beta \rightarrow \infty} 4 e^{-\beta (\xi_1 + \xi_2 + U')}$, and we obtain 
\begin{align}
G(z, \bs k) = \frac{1}{\tilde G^{-1}(z, \bs k) - (U/2)^2 \tilde G(z, \bs k)
}
\label{Eq:G1z}
\end{align}
in the orbital basis, where $\tilde G^{-1}(z, \bs k) = z  \mathbbm{1} - \qty[ \vec n(\bs k) \cdot \vec \Gamma - (\mu - U') \mathbbm{1} ]$.
Thus, the net impact of interactions in Eq. \eqref{Eq:BandBasisHamiltonian} is to shift the chemical potential $\mu \to \mu - U'$, and generate the self energy,
\begin{align}
\Sigma(z, \bs k) = (U/2)^2 \tilde G(z, \bs k).
\end{align}
The locations of poles and zeros of $G$ on the complex-$z$ plane are deduced from the roots of the denominator and numerator, respectively, of its determinant,
\begin{widetext}
\begin{align}
\det{G(z, \bs k)} = \frac{\qty[(z + (\mu - U'))^2 - |\vec n(\bs k)|^2]^2
}{\qty[
(z + (\mu - U'))^2
- \qty(\frac{U}{2} + |\vec n(\bs k)|)^2
]^2
\qty[
(z + (\mu - U'))^2
- \qty(\frac{U}{2} - |\vec n(\bs k)|)^2
]^2
}.
\label{eq:detGPsi}
\end{align}
\end{widetext}

%%%%%%%%%%%%%%%%%%%%%%%%%%%
 \begin{figure*}[!t]
\centering
\includegraphics[width=0.99\textwidth]{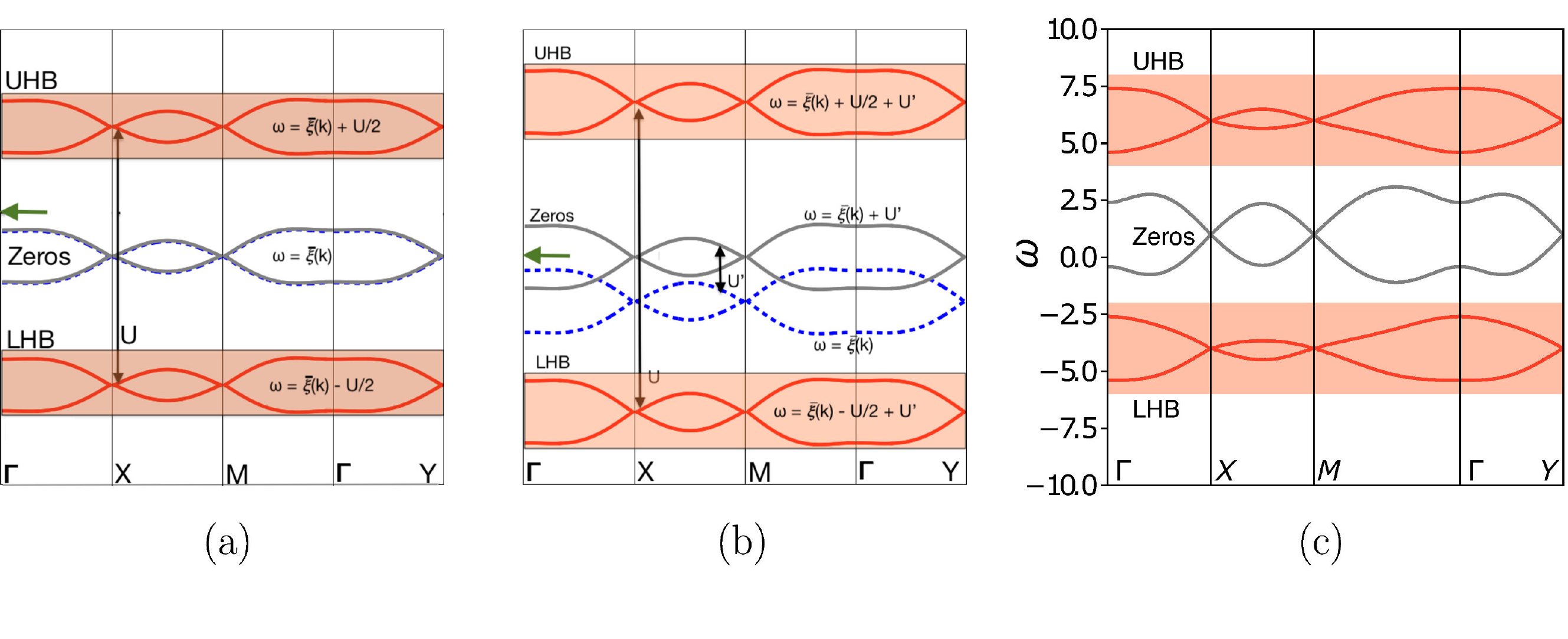}
\caption{Exact poles and zeros of the total Green's function of an interacting Dirac semimetal as $\beta \rightarrow \infty$. 
(a) The upper and lower Hubbard bands (red curves marked UHB and LHB, respectively) and zeros (gray) obtained from analytical diagonalization of the interacting Hamiltonian in the $t_{SO}/t \ll 1$ limit at $U' = 0$. 
The blue dotted curve is the original non-interacting band structure which transforms to zeros at sufficiently strong $U$. The green arrow marks the chemical potential.
(b) Same as (a) but for $U' > 0$. Here the contours of zeros and poles are shifted by $U'$. 
(c) The spectrum of the Green's function obtained by numerical exact diagonalization of the interacting Hamiltonian with $(t, t_{SO}, \mu)= (0.7, 0.42, 5)$ and $(U, U')=(10, 0.5)$.
}
\label{Fig:SquareNetUp0}
 \end{figure*}
%%%%%%%%%%%%%%%%%%%%%%%%%%%

\paragraph*{{\bf Green's function poles and zeros:}}
When $U'=0$, we have two decoupled copies of Dirac bands with only intraband interaction $U$. When both bands are below the chemical potential, we can treat the two bands separately and use the one band formula  (Eq.~(S17) in SM~\cite{sm}) for the individual bands. The partition function is simply a product of that for the individual bands and is given by $Z= \prod_{\bs k} \prod_{i = i,2} Z_{\bs k i } $ with $Z_{\bs k i} = 1+ 2 e^{-\beta \xi_i(\bs k)} + e^{-\beta\left(2 \xi_i({\bs k}) + U \right)}$. 
Each band is split into a lower and upper Hubbard band with a  crossing of zeros at the energies of the original non-interacting bands. A schematic of the various crossings is shown Fig.~\ref{Fig:SquareNetUp0}(a).
%%%
\par

The spectral function of the interacting Dirac semimetal model at $U' \neq 0$ is shown in Fig.~\ref{Fig:SquareNetUp0}(b). 
The spectral functions are analogous to the case of $U'=0$ but with the bands shifted by $U'$.  Additionally, the contour of zeros splits from the original non-interacting bands (dashed blue curves). 
The features of $G$ obtained in the limit of weak spin-orbit coupling, $(t_{SO}/t) \ll 1$, persists at more generic values of $(t_{SO}/t)$, as demonstrated by numerically diagonalizing the interacting Hamiltonian in Fig.~\ref{Fig:SquareNetUp0}(c). 

\par
%\par
\paragraph*{{\bf Symmetry constraints and spectral crossings:}}
The Green's function eigenvectors form a representation of the space group, as formulated in Ref.
\cite{Hu-Si2021} (and briefly summarized in the SM~\cite{sm}, Sec.~IV), 
and are expected to form four-fold degeneracies at the wavevectors $X$, $Y$ and $M$: As in the non-interacting case~\cite{Young-Kane2015},
the degeneracies at the $X$/$Y$ points 
are protected by $\{M_z|\frac{1}{2}\frac{1}{2} \}$ 
non-symmorphic mirror symmetry or $\{C_{2x}|\frac{1}{2} 0 \} $/$\{C_{2y}|0 \frac{1}{2} \} $ non-symmorphic screw-axis rotations,
while those at the $M$ point are also protected by
the $\{C_{2x}|\frac{1}{2} 0 \} $ and $\{C_{2y}|0 \frac{1}{2} \} $ non-symmorphic rotations. 
Our results for all three cases, with 
$t_{SO}/t \ll 1$ and $U'=0$ [Fig.~\ref{Fig:SquareNetUp0}(a)] and $U'\ne 0$ [Fig.~\ref{Fig:SquareNetUp0}(b)] as well as for the case of unconstrained ratio $t_{SO}/t$ [Fig.~\ref{Fig:SquareNetUp0}(c)], demonstrate that the entire spectra obey such degeneracies. 
This is so even though the system is a Mott insulator with a spectral gap. 
Moreover, the spectral crossing also applies to the Green's function zeros. While for the cases (a) and (b), $G(z, \bs k)$ is diagonal in the same basis that diagonalized the non-interacting Hamiltonian [see the SM~\cite{sm} (Sec.~V)], this special property does not apply to the case (c) nor to the other cases that we have analyzed
[see the SM~\cite{sm}, Fig.~2(d)(e)].
Our results, thus, illustrate our central point, namely the eigenvectors and eigenvalues of the Green's function can be used to define degeneracies by locating spectral crossings of strongly correlated systems.
{We note that in our model time-reversal symmetry protects the bands
of zeros. In its absence, Green's function zeros may appear only at
high-symmetry locations~\cite{Nagaosa2016}, but our conclusions continue to be applicable
to the poles that constitute the Hubbard bands, as discussed in details in Sec.~VI of SM~\cite{sm}. 
}

 \paragraph*{{\bf Discussion:}} Several remarks are in order. 
 First, here we  have
considered the various cases of the interactions and demonstrated the existence of both poles and zeros of the  interacting Green's function that cross at high-symmetry locations. 
These crossings are enforced by space group symmetries.
  For more generic interactions, 
  we expect an extended regime of interactions where the Green's function zeros persist and the crossings of the poles and zeros continue to be enforced by lattice symmetries. In the SM~\cite{sm}, Sec.~I, we have studied all possible symmetry allowed interaction terms for the square net lattice. We find our conclusions are robust as has been illustrated in Fig.~\ref{Fig:Schematic}.
We note that in multi-bands systems a wide variety of interactions, such as those considered here, are present that involve couplings between various internal degrees of freedom, viz. orbitals, valleys, etc.~\cite{raghu2008topological,sheng2011fractional}. 
Such terms go beyond the standard interaction in the Hubbard model, and find applicability in down-folded effective models~\cite{kang2019strong}.
For the case of zeros, we also explicitly demonstrate their symmetry-enforced crossings for Hubbard interactions via a cluster slave-spin calculation, as detailed in the Sec.~VII of the SM~\cite{sm}.
  Second, we have used both the Green's function poles and zeros to illustrate the point that the spectral crossings occur at distinct energies. 
 That the crossings of various bands of poles and zeros can be treated separately, depending on the choice of frequency,  highlights a distinct advantage of working with Green's function eigenvectors.
We have also demonstrated the symmetry enforced crossings of zeros and poles in another lattice model (diamond lattice~\cite{fu2007PRL, fu2007PRB}) with HK interaction, which can be found in Sec.~VIII in the SM~\cite{sm}.
Third, 
the kind of spectral crossings we have discussed sets the stage to analyze the form of topology
when strong correlations turn a non-interacting topological semimetals into a Mott insulator. 
{One way to define such topology is through a classification of the eigenvectors of the Greens function~\cite{Hu-Si2021}, which leads to frequency-dependent topological invariants, and associates a   monopole charge to the crossings of the bands of zeros.~\cite{setty2023topological}.
An additional outcome of this method is the elucidation of the manner in which the Green's function zeros may contribute to topological response in a Mott insulating state~\cite{setty2023electronic}.
}
 Finally, by demonstrating symmetry constraints and spectral crossing in an extreme interacting setting, our work provides support to the work of Ref~\cite{Hu-Si2021}. There, spectral crossings through symmetry constraints plays a central role in realizing topological semimetals without Landau quasiparticles.
\par

\paragraph*{{\bf Implications for experiments and materials:}}
The Green's function eigenvector formulation
\cite{Hu-Si2021} is expected to yield 
spectral crossings in other symmetry settings. For example, in the case of Bi$_2$CuO$_4$ an 
eight-fold degeneracy\cite{Wieder2016} 
is expected at certain high symmetry wavevectors in its non-interacting bandstructure
\cite{Bradlyn2016}. 
The system is in fact strongly interacting
\cite{DiSante2017,goldoni1994electronic}
and we expect that its paramagnetic Mott insulator state (above its N\'{e}el temperature of $50$ K \cite{garcia1990crystal}) will feature eight-fold spectral crossings in the form we have described in some detail here. Probing the spectral function by applying a symmetry-breaking perturbation will allow 
for experimentally revealing this spectral crossing.

Separately, proximity 
to an orbital-selective Mott insulating state has 
recently been advanced as a means of generating 
Kondo-driven topological semimetals 
in $d$-electron-based systems that host topological flat bands \cite{Chen-emergent2022,Hu-FB2022,Che23.1x}. We can expect that the type of spectral crossings 
of both the peaks and zeros as discussed here may play an important role in such orbital-selective Mott states.
\par

  To conclude, we provide a proof-of-principle demonstration of how lattice symmetries can be used to constrain excitations in an extreme limit of strongly correlated systems. Typically in non-interacting systems, eigenstates of the Hamiltonian and their symmetry operators are used to indicate topology by diagnosing conditions for symmetry protected band crossings.  For interacting systems, we recently showed that Green's function eigenvectors form a representation of the lattice space group and can be used to
diagnose and realize topology~\cite{Hu-Si2021}.
  Here we study an exactly solvable model of a Mott insulator where eigenvectors and eigenvalues of the Green's function can be used to locate crossings of poles and zeros in momentum space. Together with the realization of topological semimetals without Landau quasiparticles, our work demonstrates
the power that the Green's function formulation of symmetry constraints displays for realizing non-trivial spectral crossings in fully interacting settings. We can 
expect the approach to be important for the
symmetry-based design of topological quantum materials.

\acknowledgements{
We thank N. Nagaosa for a useful discussion.
Work at Rice has primarily been supported 
by the National Science Foundation
under Grant No. DMR-2220603 (F.X. and Q.S.),
by the Air Force Office of Scientific Research under Grant No.
FA9550-21-1-0356 (C.S. and S.S.), 
and by the Robert A. Welch Foundation Grant No. C-1411 (L.C.)
and the Vannevar Bush Faculty Fellowship ONR-VB N00014-23-1-2870 (Q.S.). The
majority of the computational calculations have been performed on the Shared University Grid
at Rice funded by NSF under Grant EIA-0216467, a partnership between Rice University, Sun
Microsystems, and Sigma Solutions, Inc., the Big-Data Private-Cloud Research Cyberinfrastructure
MRI-award funded by NSF under Grant No. CNS-1338099, and the Extreme Science and
Engineering Discovery Environment (XSEDE) by NSF under Grant No. DMR170109. H.H. acknowledges
the support of the European Research Council (ERC) under the European Union's 
Horizon 2020 research and innovation program (Grant Agreement No.\ 101020833). Work in
Vienna was supported by the Austrian Science Fund (project I 5868-N - FOR 5249 - QUAST)
and the ERC (Advanced Grant CorMeTop, No.\ 101055088). J.C. acknowledges the support of
the National Science Foundation under Grant No. DMR-1942447, support from the Alfred P.
Sloan Foundation through a Sloan Research Fellowship and the support of the Flatiron Institute,
a division of the Simons Foundation. S.P. and Q.S. acknowledge the hospitality of the Aspen Center for
Physics, which is supported by NSF grant No. 
PHY-2210452.

\textit{Note added:} After the completion of this manuscript, a recent work addressing a different model with a focus on the 
  Green's function zeros and poles in interacting topological insulators became available 
(N. Wagner et al.\cite{wagner2023mott}).
}
  
 \!\!\!\!\!\!\!\!$^\oplus$ csetty@rice.edu \\
 $^\dagger$ shouvik.sur@rice.edu \\
 $^*$ These authors contributed equally
\bibliography{Zeros2.bib}

\begin{thebibliography}{54}%
\makeatletter
\providecommand \@ifxundefined [1]{%
 \@ifx{#1\undefined}
}%
\providecommand \@ifnum [1]{%
 \ifnum #1\expandafter \@firstoftwo
 \else \expandafter \@secondoftwo
 \fi
}%
\providecommand \@ifx [1]{%
 \ifx #1\expandafter \@firstoftwo
 \else \expandafter \@secondoftwo
 \fi
}%
\providecommand \natexlab [1]{#1}%
\providecommand \enquote  [1]{``#1''}%
\providecommand \bibnamefont  [1]{#1}%
\providecommand \bibfnamefont [1]{#1}%
\providecommand \citenamefont [1]{#1}%
\providecommand \href@noop [0]{\@secondoftwo}%
\providecommand \href [0]{\begingroup \@sanitize@url \@href}%
\providecommand \@href[1]{\@@startlink{#1}\@@href}%
\providecommand \@@href[1]{\endgroup#1\@@endlink}%
\providecommand \@sanitize@url [0]{\catcode `\\12\catcode `\$12\catcode
  `\&12\catcode `\#12\catcode `\^12\catcode `\_12\catcode `\%12\relax}%
\providecommand \@@startlink[1]{}%
\providecommand \@@endlink[0]{}%
\providecommand \url  [0]{\begingroup\@sanitize@url \@url }%
\providecommand \@url [1]{\endgroup\@href {#1}{\urlprefix }}%
\providecommand \urlprefix  [0]{URL }%
\providecommand \Eprint [0]{\href }%
\providecommand \doibase [0]{http://dx.doi.org/}%
\providecommand \selectlanguage [0]{\@gobble}%
\providecommand \bibinfo  [0]{\@secondoftwo}%
\providecommand \bibfield  [0]{\@secondoftwo}%
\providecommand \translation [1]{[#1]}%
\providecommand \BibitemOpen [0]{}%
\providecommand \bibitemStop [0]{}%
\providecommand \bibitemNoStop [0]{.\EOS\space}%
\providecommand \EOS [0]{\spacefactor3000\relax}%
\providecommand \BibitemShut  [1]{\csname bibitem#1\endcsname}%
\let\auto@bib@innerbib\@empty
%</preamble>
\bibitem [{\citenamefont {Armitage}\ \emph {et~al.}(2018)\citenamefont
  {Armitage}, \citenamefont {Mele},\ and\ \citenamefont
  {Vishwanath}}]{Armitage2017}%
  \BibitemOpen
  \bibfield  {author} {\bibinfo {author} {\bibfnamefont {N.~P.}\ \bibnamefont
  {Armitage}}, \bibinfo {author} {\bibfnamefont {E.~J.}\ \bibnamefont {Mele}},
  \ and\ \bibinfo {author} {\bibfnamefont {Ashvin}\ \bibnamefont
  {Vishwanath}},\ }\bibfield  {title} {\enquote {\bibinfo {title} {{Weyl} and
  {Dirac} semimetals in three-dimensional solids},}\ }\href@noop {} {\bibfield  {journal} {\bibinfo  {journal}
  {Rev. Mod. Phys.}\ }\textbf {\bibinfo {volume} {90}},\ \bibinfo {pages}
  {015001} (\bibinfo {year} {2018})}\BibitemShut {NoStop}%
\bibitem [{\citenamefont {Nagaosa}\ \emph {et~al.}(2020)\citenamefont
  {Nagaosa}, \citenamefont {Morimoto},\ and\ \citenamefont
  {Tokura}}]{Nagaosa2020}%
  \BibitemOpen
  \bibfield  {author} {\bibinfo {author} {\bibfnamefont {Naoto}\ \bibnamefont
  {Nagaosa}}, \bibinfo {author} {\bibfnamefont {Takahiro}\ \bibnamefont
  {Morimoto}}, \ and\ \bibinfo {author} {\bibfnamefont {Yoshinori}\
  \bibnamefont {Tokura}},\ }\bibfield  {title} {\enquote {\bibinfo {title}
  {Transport, magnetic and optical properties of {W}eyl materials},}\ }\href@noop {} {\bibfield  {journal} {\bibinfo
  {journal} {Nature Reviews Materials}\ }\textbf {\bibinfo {volume} {5}},\
  \bibinfo {pages} {621--636} (\bibinfo {year} {2020})}\BibitemShut {NoStop}%
\bibitem [{\citenamefont {Bradlyn}\ \emph {et~al.}(2017)\citenamefont
  {Bradlyn}, \citenamefont {Elcoro}, \citenamefont {Cano}, \citenamefont
  {Vergniory}, \citenamefont {Wang}, \citenamefont {Felser}, \citenamefont
  {Aroyo},\ and\ \citenamefont {Bernevig}}]{Bradlyn2017}%
  \BibitemOpen
  \bibfield  {author} {\bibinfo {author} {\bibfnamefont {Barry}\ \bibnamefont
  {Bradlyn}}, \bibinfo {author} {\bibfnamefont {L.}~\bibnamefont {Elcoro}},
  \bibinfo {author} {\bibfnamefont {Jennifer}\ \bibnamefont {Cano}}, \bibinfo
  {author} {\bibfnamefont {M.~G.}\ \bibnamefont {Vergniory}}, \bibinfo {author}
  {\bibfnamefont {Zhijun}\ \bibnamefont {Wang}}, \bibinfo {author}
  {\bibfnamefont {C.}~\bibnamefont {Felser}}, \bibinfo {author} {\bibfnamefont
  {M.~I.}\ \bibnamefont {Aroyo}}, \ and\ \bibinfo {author} {\bibfnamefont
  {B.~Andrei}\ \bibnamefont {Bernevig}},\ }\bibfield  {title} {\enquote
  {\bibinfo {title} {Topological quantum chemistry},}\ } {\bibfield  {journal} {\bibinfo  {journal} {Nature}\
  }\textbf {\bibinfo {volume} {547}},\ \bibinfo {pages} {298--305} (\bibinfo
  {year} {2017})}\BibitemShut {NoStop}%
\bibitem [{\citenamefont {Cano}\ \emph {et~al.}(2018)\citenamefont {Cano},
  \citenamefont {Bradlyn}, \citenamefont {Wang}, \citenamefont {Elcoro},
  \citenamefont {Vergniory}, \citenamefont {Felser}, \citenamefont {Aroyo},\
  and\ \citenamefont {Bernevig}}]{Cano2018}%
  \BibitemOpen
  \bibfield  {author} {\bibinfo {author} {\bibfnamefont {Jennifer}\
  \bibnamefont {Cano}}, \bibinfo {author} {\bibfnamefont {Barry}\ \bibnamefont
  {Bradlyn}}, \bibinfo {author} {\bibfnamefont {Zhijun}\ \bibnamefont {Wang}},
  \bibinfo {author} {\bibfnamefont {L.}~\bibnamefont {Elcoro}}, \bibinfo
  {author} {\bibfnamefont {M.~G.}\ \bibnamefont {Vergniory}}, \bibinfo {author}
  {\bibfnamefont {C.}~\bibnamefont {Felser}}, \bibinfo {author} {\bibfnamefont
  {M.~I.}\ \bibnamefont {Aroyo}}, \ and\ \bibinfo {author} {\bibfnamefont
  {B.~Andrei}\ \bibnamefont {Bernevig}},\ }\bibfield  {title} {\enquote
  {\bibinfo {title} {Building blocks of topological quantum chemistry:
  Elementary band representations},}\ } {\bibfield  {journal} {\bibinfo  {journal} {Phys.
  Rev. B}\ }\textbf {\bibinfo {volume} {97}},\ \bibinfo {pages} {035139}
  (\bibinfo {year} {2018})}\BibitemShut {NoStop}%
\bibitem [{\citenamefont {Po}\ \emph {et~al.}(2017)\citenamefont {Po},
  \citenamefont {Vishwanath},\ and\ \citenamefont {Watanabe}}]{Po2017}%
  \BibitemOpen
  \bibfield  {author} {\bibinfo {author} {\bibfnamefont {Hoi~Chun}\
  \bibnamefont {Po}}, \bibinfo {author} {\bibfnamefont {Ashvin}\ \bibnamefont
  {Vishwanath}}, \ and\ \bibinfo {author} {\bibfnamefont {Haruki}\ \bibnamefont
  {Watanabe}},\ }\bibfield  {title} {\enquote {\bibinfo {title} {Symmetry-based
  indicators of band topology in the 230 space groups},}\ }\href@noop {} {\bibfield  {journal} {\bibinfo  {journal}
  {Nature Communications}\ }\textbf {\bibinfo {volume} {8}},\ \bibinfo {pages}
  {50} (\bibinfo {year} {2017})}\BibitemShut {NoStop}%
\bibitem [{\citenamefont {Watanabe}\ \emph {et~al.}(2016)\citenamefont
  {Watanabe}, \citenamefont {Po}, \citenamefont {Zaletel},\ and\ \citenamefont
  {Vishwanath}}]{Watanabe2017}%
  \BibitemOpen
  \bibfield  {author} {\bibinfo {author} {\bibfnamefont {Haruki}\ \bibnamefont
  {Watanabe}}, \bibinfo {author} {\bibfnamefont {Hoi~Chun}\ \bibnamefont {Po}},
  \bibinfo {author} {\bibfnamefont {Michael~P.}\ \bibnamefont {Zaletel}}, \
  and\ \bibinfo {author} {\bibfnamefont {Ashvin}\ \bibnamefont {Vishwanath}},\
  }\bibfield  {title} {\enquote {\bibinfo {title} {Filling-enforced gaplessness
  in band structures of the 230 space groups},}\ }\href@noop {} {\bibfield  {journal} {\bibinfo  {journal}
  {Phys. Rev. Lett.}\ }\textbf {\bibinfo {volume} {117}},\ \bibinfo {pages}
  {096404} (\bibinfo {year} {2016})}\BibitemShut {NoStop}%
\bibitem [{\citenamefont {Cano}\ and\ \citenamefont
  {Bradlyn}(2021)}]{cano2021band}%
  \BibitemOpen
  \bibfield  {author} {\bibinfo {author} {\bibfnamefont {Jennifer}\
  \bibnamefont {Cano}}\ and\ \bibinfo {author} {\bibfnamefont {Barry}\
  \bibnamefont {Bradlyn}},\ }\bibfield  {title} {\enquote {\bibinfo {title}
  {Band representations and topological quantum chemistry},}\ }\href@noop {}
  {\bibfield  {journal} {\bibinfo  {journal} {Annu.\ Rev.\ Condens.\ Matter
  Phys.}\ }\textbf {\bibinfo {volume} {12}},\ \bibinfo {pages} {225--246}
  (\bibinfo {year} {2021})}\BibitemShut {NoStop}%
\bibitem [{\citenamefont {Son}(2007)}]{son2007}%
  \BibitemOpen
  \bibfield  {author} {\bibinfo {author} {\bibfnamefont {DT}~\bibnamefont
  {Son}},\ }\bibfield  {title} {\enquote {\bibinfo {title} {Quantum critical
  point in graphene approached in the limit of infinitely strong coulomb
  interaction},}\ }\href@noop {} {\bibfield
  {journal} {\bibinfo  {journal} {Phys. Rev. B}\ }\textbf {\bibinfo {volume}
  {75}},\ \bibinfo {pages} {235423} (\bibinfo {year} {2007})}\BibitemShut
  {NoStop}%
\bibitem [{\citenamefont {Sun}\ \emph {et~al.}(2009)\citenamefont {Sun},
  \citenamefont {Yao}, \citenamefont {Fradkin},\ and\ \citenamefont
  {Kivelson}}]{sun2009}%
  \BibitemOpen
  \bibfield  {author} {\bibinfo {author} {\bibfnamefont {Kai}\ \bibnamefont
  {Sun}}, \bibinfo {author} {\bibfnamefont {Hong}\ \bibnamefont {Yao}},
  \bibinfo {author} {\bibfnamefont {Eduardo}\ \bibnamefont {Fradkin}}, \ and\
  \bibinfo {author} {\bibfnamefont {Steven~A}\ \bibnamefont {Kivelson}},\
  }\bibfield  {title} {\enquote {\bibinfo {title} {Topological insulators and
  nematic phases from spontaneous symmetry breaking in 2d {F}ermi systems with
  a quadratic band crossing},}\ }\href@noop {} {\bibfield  {journal} {\bibinfo
  {journal} {Phys. Rev. Lett.}\ }\textbf {\bibinfo {volume} {103}},\ \bibinfo
  {pages} {046811} (\bibinfo {year} {2009})}\BibitemShut {NoStop}%
\bibitem [{\citenamefont {Vafek}\ and\ \citenamefont {Yang}(2010)}]{vafek2010}%
  \BibitemOpen
  \bibfield  {author} {\bibinfo {author} {\bibfnamefont {Oskar}\ \bibnamefont
  {Vafek}}\ and\ \bibinfo {author} {\bibfnamefont {Kun}\ \bibnamefont {Yang}},\
  }\bibfield  {title} {\enquote {\bibinfo {title} {Many-body instability of
  coulomb interacting bilayer graphene: Renormalization group approach},}\
  }\href@noop {} {\bibfield  {journal} {\bibinfo
  {journal} {Phys. Rev. B}\ }\textbf {\bibinfo {volume} {81}},\ \bibinfo
  {pages} {041401} (\bibinfo {year} {2010})}\BibitemShut {NoStop}%
\bibitem [{\citenamefont {Herbut}\ \emph {et~al.}(2009)\citenamefont {Herbut},
  \citenamefont {Juri{\v{c}}i{\'c}},\ and\ \citenamefont {Roy}}]{herbut2009}%
  \BibitemOpen
  \bibfield  {author} {\bibinfo {author} {\bibfnamefont {Igor~F}\ \bibnamefont
  {Herbut}}, \bibinfo {author} {\bibfnamefont {Vladimir}\ \bibnamefont
  {Juri{\v{c}}i{\'c}}}, \ and\ \bibinfo {author} {\bibfnamefont {Bitan}\
  \bibnamefont {Roy}},\ }\bibfield  {title} {\enquote {\bibinfo {title} {Theory
  of interacting electrons on the honeycomb lattice},}\ }\href@noop {} {\bibfield  {journal} {\bibinfo  {journal} {Phys.
  Rev. B}\ }\textbf {\bibinfo {volume} {79}},\ \bibinfo {pages} {085116}
  (\bibinfo {year} {2009})}\BibitemShut {NoStop}%
\bibitem [{\citenamefont {Sur}\ and\ \citenamefont
  {Nandkishore}(2016)}]{sur2016}%
  \BibitemOpen
  \bibfield  {author} {\bibinfo {author} {\bibfnamefont {Shouvik}\ \bibnamefont
  {Sur}}\ and\ \bibinfo {author} {\bibfnamefont {Rahul}\ \bibnamefont
  {Nandkishore}},\ }\bibfield  {title} {\enquote {\bibinfo {title}
  {Instabilities of {W}eyl loop semimetals},}\ }\href@noop {} {\bibfield
  {journal} {\bibinfo  {journal} {New Journal of Physics}\ }\textbf {\bibinfo
  {volume} {18}},\ \bibinfo {pages} {115006} (\bibinfo {year}
  {2016})}\BibitemShut {NoStop}%
\bibitem [{\citenamefont {Huh}\ \emph {et~al.}(2016)\citenamefont {Huh},
  \citenamefont {Moon},\ and\ \citenamefont {Kim}}]{huh2016}%
  \BibitemOpen
  \bibfield  {author} {\bibinfo {author} {\bibfnamefont {Yejin}\ \bibnamefont
  {Huh}}, \bibinfo {author} {\bibfnamefont {Eun-Gook}\ \bibnamefont {Moon}}, \
  and\ \bibinfo {author} {\bibfnamefont {Yong~Baek}\ \bibnamefont {Kim}},\
  }\bibfield  {title} {\enquote {\bibinfo {title} {Long-range coulomb
  interaction in nodal-ring semimetals},}\ }\href@noop {} {\bibfield  {journal}
  {\bibinfo  {journal} {Phys. Rev. B}\ }\textbf {\bibinfo {volume} {93}},\
  \bibinfo {pages} {035138} (\bibinfo {year} {2016})}\BibitemShut {NoStop}%
\bibitem [{\citenamefont {Roy}(2017)}]{roy2017}%
  \BibitemOpen
  \bibfield  {author} {\bibinfo {author} {\bibfnamefont {Bitan}\ \bibnamefont
  {Roy}},\ }\bibfield  {title} {\enquote {\bibinfo {title} {Interacting
  nodal-line semimetal: Proximity effect and spontaneous symmetry breaking},}\
  }\href@noop {} {\bibfield  {journal} {\bibinfo
  {journal} {Phys. Rev. B}\ }\textbf {\bibinfo {volume} {96}},\ \bibinfo
  {pages} {041113} (\bibinfo {year} {2017})}\BibitemShut {NoStop}%
\bibitem [{\citenamefont {Sur}\ and\ \citenamefont {Roy}(2019)}]{sur2019}%
  \BibitemOpen
  \bibfield  {author} {\bibinfo {author} {\bibfnamefont {Shouvik}\ \bibnamefont
  {Sur}}\ and\ \bibinfo {author} {\bibfnamefont {Bitan}\ \bibnamefont {Roy}},\
  }\bibfield  {title} {\enquote {\bibinfo {title} {Unifying interacting nodal
  semimetals: A new route to strong coupling},}\ }\href@noop {} {\bibfield  {journal} {\bibinfo  {journal}
  {Phys. Rev. Lett.}\ }\textbf {\bibinfo {volume} {123}},\ \bibinfo {pages}
  {207601} (\bibinfo {year} {2019})}\BibitemShut {NoStop}%
\bibitem [{\citenamefont {Wang}\ and\ \citenamefont
  {Zhang}(2012)}]{Wang-Zhang_PRX2012}%
  \BibitemOpen
  \bibfield  {author} {\bibinfo {author} {\bibfnamefont {Zhong}\ \bibnamefont
  {Wang}}\ and\ \bibinfo {author} {\bibfnamefont {Shou-Cheng}\ \bibnamefont
  {Zhang}},\ }\bibfield  {title} {\enquote {\bibinfo {title} {Simplified
  topological invariants for interacting insulators},}\ }\href@noop {} {\bibfield  {journal} {\bibinfo  {journal} {Phys.
  Rev. X}\ }\textbf {\bibinfo {volume} {2}},\ \bibinfo {pages} {031008}
  (\bibinfo {year} {2012})}\BibitemShut {NoStop}%
\bibitem [{\citenamefont {Wang}\ and\ \citenamefont
  {Yan}(2013)}]{Wang-Yan2013}%
  \BibitemOpen
  \bibfield  {author} {\bibinfo {author} {\bibfnamefont {Zhong}\ \bibnamefont
  {Wang}}\ and\ \bibinfo {author} {\bibfnamefont {Binghai}\ \bibnamefont
  {Yan}},\ }\bibfield  {title} {\enquote {\bibinfo {title} {Topological
  hamiltonian as an exact tool for topological invariants},}\ }\href@noop {} {\bibfield  {journal} {\bibinfo  {journal}
  {Journal of Physics: Condensed Matter}\ }\textbf {\bibinfo {volume} {25}},\
  \bibinfo {pages} {155601} (\bibinfo {year} {2013})}\BibitemShut {NoStop}%
\bibitem [{\citenamefont {Lessnich}\ \emph {et~al.}(2021)\citenamefont
  {Lessnich}, \citenamefont {Winter}, \citenamefont {Iraola}, \citenamefont
  {Vergniory},\ and\ \citenamefont {Valent\'{\i}}}]{Lessnich2021}%
  \BibitemOpen
  \bibfield  {author} {\bibinfo {author} {\bibfnamefont {Dominik}\ \bibnamefont
  {Lessnich}}, \bibinfo {author} {\bibfnamefont {Stephen~M.}\ \bibnamefont
  {Winter}}, \bibinfo {author} {\bibfnamefont {Mikel}\ \bibnamefont {Iraola}},
  \bibinfo {author} {\bibfnamefont {Maia~G.}\ \bibnamefont {Vergniory}}, \ and\
  \bibinfo {author} {\bibfnamefont {Roser}\ \bibnamefont {Valent\'{\i}}},\
  }\bibfield  {title} {\enquote {\bibinfo {title} {Elementary band
  representations for the single-particle {G}reen's function of interacting
  topological insulators},}\ }\href@noop {}
  {\bibfield  {journal} {\bibinfo  {journal} {Phys. Rev. B}\ }\textbf {\bibinfo
  {volume} {104}},\ \bibinfo {pages} {085116} (\bibinfo {year}
  {2021})}\BibitemShut {NoStop}%
\bibitem [{\citenamefont {Soldini}\ \emph {et~al.}(2023)\citenamefont
  {Soldini}, \citenamefont {Astrakhantsev}, \citenamefont {Iraola},
  \citenamefont {Tiwari}, \citenamefont {Fischer}, \citenamefont
  {Valent{\'\i}}, \citenamefont {Vergniory}, \citenamefont {Wagner},\ and\
  \citenamefont {Neupert}}]{Soldini2022}%
  \BibitemOpen
  \bibfield  {author} {\bibinfo {author} {\bibfnamefont {Martina~O}\
  \bibnamefont {Soldini}}, \bibinfo {author} {\bibfnamefont {Nikita}\
  \bibnamefont {Astrakhantsev}}, \bibinfo {author} {\bibfnamefont {Mikel}\
  \bibnamefont {Iraola}}, \bibinfo {author} {\bibfnamefont {Apoorv}\
  \bibnamefont {Tiwari}}, \bibinfo {author} {\bibfnamefont {Mark~H}\
  \bibnamefont {Fischer}}, \bibinfo {author} {\bibfnamefont {Roser}\
  \bibnamefont {Valent{\'\i}}}, \bibinfo {author} {\bibfnamefont {Maia~G}\
  \bibnamefont {Vergniory}}, \bibinfo {author} {\bibfnamefont {Glenn}\
  \bibnamefont {Wagner}}, \ and\ \bibinfo {author} {\bibfnamefont {Titus}\
  \bibnamefont {Neupert}},\ }\bibfield  {title} {\enquote {\bibinfo {title}
  {Interacting topological quantum chemistry of {M}ott atomic limits},}\
  }\href@noop {} {\bibfield  {journal} {\bibinfo  {journal} {Phys. Rev. B}\
  }\textbf {\bibinfo {volume} {107}},\ \bibinfo {pages} {245145} (\bibinfo
  {year} {2023})}\BibitemShut {NoStop}%
\bibitem [{\citenamefont {Hu}\ \emph {et~al.}(2021)\citenamefont {Hu},
  \citenamefont {Chen}, \citenamefont {Setty}, \citenamefont {Garcia-Diez},
  \citenamefont {Grefe}, \citenamefont {Yan}, \citenamefont {Lu\v{z}nik},
  \citenamefont {Reumann}, \citenamefont {Prokofiev}, \citenamefont {Kirchner},
  \citenamefont {Vergniory}, \citenamefont {Paschen}, \citenamefont {Cano},\
  and\ \citenamefont {Si}}]{Hu-Si2021}%
  \BibitemOpen
  \bibfield  {author} {\bibinfo {author} {\bibfnamefont {Haoyu~Hu}\
  \bibnamefont {Hu}}, \bibinfo {author} {\bibfnamefont {Lei}\ \bibnamefont
  {Chen}}, \bibinfo {author} {\bibfnamefont {Chandan}\ \bibnamefont {Setty}},
  \bibinfo {author} {\bibfnamefont {Mikel}\ \bibnamefont {Garcia-Diez}},
  \bibinfo {author} {\bibfnamefont {Sarah~E}\ \bibnamefont {Grefe}}, \bibinfo
  {author} {\bibfnamefont {Xinlin}\ \bibnamefont {Yan}}, \bibinfo {author}
  {\bibfnamefont {Monika}\ \bibnamefont {Lu\v{z}nik}}, \bibinfo {author}
  {\bibfnamefont {Nikolas}\ \bibnamefont {Reumann}}, \bibinfo {author}
  {\bibfnamefont {Andrey}\ \bibnamefont {Prokofiev}}, \bibinfo {author}
  {\bibfnamefont {Stefan}\ \bibnamefont {Kirchner}}, \bibinfo {author}
  {\bibfnamefont {Maia~G.}\ \bibnamefont {Vergniory}}, \bibinfo {author}
  {\bibfnamefont {Silke}\ \bibnamefont {Paschen}}, \bibinfo {author}
  {\bibfnamefont {Jennifer}\ \bibnamefont {Cano}}, \ and\ \bibinfo {author}
  {\bibfnamefont {Qimiao}\ \bibnamefont {Si}},\ }\bibfield  {title} {\enquote
  {\bibinfo {title} {Topological semimetals without quasiparticles},}\
  }\href@noop {} {\bibfield  {journal} {\bibinfo  {journal} {arXiv preprint
  arXiv:2110.06182}\ } (\bibinfo {year} {2021})}\BibitemShut {NoStop}%
\bibitem [{\citenamefont {Chen}\ \emph
  {et~al.}(2022{\natexlab{a}})\citenamefont {Chen}, \citenamefont {Setty},
  \citenamefont {Hu}, \citenamefont {Vergniory}, \citenamefont {Grefe},
  \citenamefont {Fischer}, \citenamefont {Yan}, \citenamefont {Eguchi},
  \citenamefont {Prokofiev}, \citenamefont {Paschen} \emph
  {et~al.}}]{Chen-Si2022}%
  \BibitemOpen
  \bibfield  {author} {\bibinfo {author} {\bibfnamefont {Lei}\ \bibnamefont
  {Chen}}, \bibinfo {author} {\bibfnamefont {Chandan}\ \bibnamefont {Setty}},
  \bibinfo {author} {\bibfnamefont {Haoyu}\ \bibnamefont {Hu}}, \bibinfo
  {author} {\bibfnamefont {Maia~G}\ \bibnamefont {Vergniory}}, \bibinfo
  {author} {\bibfnamefont {Sarah~E}\ \bibnamefont {Grefe}}, \bibinfo {author}
  {\bibfnamefont {Lukas}\ \bibnamefont {Fischer}}, \bibinfo {author}
  {\bibfnamefont {Xinlin}\ \bibnamefont {Yan}}, \bibinfo {author}
  {\bibfnamefont {Gaku}\ \bibnamefont {Eguchi}}, \bibinfo {author}
  {\bibfnamefont {Andrey}\ \bibnamefont {Prokofiev}}, \bibinfo {author}
  {\bibfnamefont {Silke}\ \bibnamefont {Paschen}},  \emph {et~al.},\ }\bibfield
   {title} {\enquote {\bibinfo {title} {Topological semimetal driven by strong
  correlations and crystalline symmetry},}\ }\href@noop {} {\bibfield
  {journal} {\bibinfo  {journal} {Nature Physics}\ }\textbf {\bibinfo {volume}
  {18}},\ \bibinfo {pages} {1341--1346} (\bibinfo {year}
  {2022}{\natexlab{a}})}\BibitemShut {NoStop}%
\bibitem [{\citenamefont {Lai}\ \emph {et~al.}(2018)\citenamefont {Lai},
  \citenamefont {Grefe}, \citenamefont {Paschen},\ and\ \citenamefont
  {Si}}]{Lai2018}%
  \BibitemOpen
  \bibfield  {author} {\bibinfo {author} {\bibfnamefont {Hsin-Hua}\
  \bibnamefont {Lai}}, \bibinfo {author} {\bibfnamefont {Sarah~E.}\
  \bibnamefont {Grefe}}, \bibinfo {author} {\bibfnamefont {Silke}\ \bibnamefont
  {Paschen}}, \ and\ \bibinfo {author} {\bibfnamefont {Qimiao}\ \bibnamefont
  {Si}},\ }\bibfield  {title} {\enquote {\bibinfo {title} {Weyl-{K}ondo
  semimetal in heavy-{F}ermion systems},}\ }\href@noop {} {\bibfield  {journal} {\bibinfo  {journal} {Proc.
  Natl. Acad. Sci. U.S.A.}\ }\textbf {\bibinfo {volume} {115}},\ \bibinfo
  {pages} {93} (\bibinfo {year} {2018})}\BibitemShut {NoStop}%
\bibitem [{\citenamefont {Dzsaber}\ \emph {et~al.}(2017)\citenamefont
  {Dzsaber}, \citenamefont {Prochaska}, \citenamefont {Sidorenko},
  \citenamefont {Eguchi}, \citenamefont {Svagera}, \citenamefont {Waas},
  \citenamefont {Prokofiev}, \citenamefont {Si},\ and\ \citenamefont
  {Paschen}}]{Dzsaber2017}%
  \BibitemOpen
  \bibfield  {author} {\bibinfo {author} {\bibfnamefont {S.}~\bibnamefont
  {Dzsaber}}, \bibinfo {author} {\bibfnamefont {L.}~\bibnamefont {Prochaska}},
  \bibinfo {author} {\bibfnamefont {A.}~\bibnamefont {Sidorenko}}, \bibinfo
  {author} {\bibfnamefont {G.}~\bibnamefont {Eguchi}}, \bibinfo {author}
  {\bibfnamefont {R.}~\bibnamefont {Svagera}}, \bibinfo {author} {\bibfnamefont
  {M.}~\bibnamefont {Waas}}, \bibinfo {author} {\bibfnamefont {A.}~\bibnamefont
  {Prokofiev}}, \bibinfo {author} {\bibfnamefont {Q.}~\bibnamefont {Si}}, \
  and\ \bibinfo {author} {\bibfnamefont {S.}~\bibnamefont {Paschen}},\
  }\bibfield  {title} {\enquote {\bibinfo {title} {Kondo insulator to semimetal
  transformation tuned by spin-orbit coupling},}\ }\href@noop {} {\bibfield  {journal} {\bibinfo  {journal}
  {Phys. Rev. Lett.}\ }\textbf {\bibinfo {volume} {118}},\ \bibinfo {pages}
  {246601} (\bibinfo {year} {2017})}\BibitemShut {NoStop}%
\bibitem [{\citenamefont {Dzsaber}\ \emph {et~al.}(2021)\citenamefont
  {Dzsaber}, \citenamefont {Yan}, \citenamefont {Taupin}, \citenamefont
  {Eguchi}, \citenamefont {Prokofiev}, \citenamefont {Shiroka}, \citenamefont
  {Blaha}, \citenamefont {Rubel}, \citenamefont {Grefe}, \citenamefont {Lai},
  \citenamefont {Si},\ and\ \citenamefont {Paschen}}]{Dzs-giant21.1}%
  \BibitemOpen
  \bibfield  {author} {\bibinfo {author} {\bibfnamefont {Sami}\ \bibnamefont
  {Dzsaber}}, \bibinfo {author} {\bibfnamefont {Xinlin}\ \bibnamefont {Yan}},
  \bibinfo {author} {\bibfnamefont {Mathieu}\ \bibnamefont {Taupin}}, \bibinfo
  {author} {\bibfnamefont {Gaku}\ \bibnamefont {Eguchi}}, \bibinfo {author}
  {\bibfnamefont {Andrey}\ \bibnamefont {Prokofiev}}, \bibinfo {author}
  {\bibfnamefont {Toni}\ \bibnamefont {Shiroka}}, \bibinfo {author}
  {\bibfnamefont {Peter}\ \bibnamefont {Blaha}}, \bibinfo {author}
  {\bibfnamefont {Oleg}\ \bibnamefont {Rubel}}, \bibinfo {author}
  {\bibfnamefont {Sarah~E.}\ \bibnamefont {Grefe}}, \bibinfo {author}
  {\bibfnamefont {Hsin-Hua}\ \bibnamefont {Lai}}, \bibinfo {author}
  {\bibfnamefont {Qimiao}\ \bibnamefont {Si}}, \ and\ \bibinfo {author}
  {\bibfnamefont {Silke}\ \bibnamefont {Paschen}},\ }\bibfield  {title}
  {\enquote {\bibinfo {title} {Giant spontaneous hall effect in a nonmagnetic
  {W}eyl {K}ondo semimetal},}\ }\href@noop {}
  {\bibfield  {journal} {\bibinfo  {journal} {PNAS}\ }\textbf {\bibinfo
  {volume} {118}},\ \bibinfo {pages} {e2013386118} (\bibinfo {year}
  {2021})}\BibitemShut {NoStop}%
\bibitem [{\citenamefont {Morimoto}\ and\ \citenamefont
  {Nagaosa}(2016)}]{Nagaosa2016}%
  \BibitemOpen
  \bibfield  {author} {\bibinfo {author} {\bibfnamefont {Takahiro}\
  \bibnamefont {Morimoto}}\ and\ \bibinfo {author} {\bibfnamefont {Naoto}\
  \bibnamefont {Nagaosa}},\ }\bibfield  {title} {\enquote {\bibinfo {title}
  {Weyl {M}ott insulator},}\ }\href@noop {} {\bibfield  {journal} {\bibinfo
  {journal} {Scientific reports}\ }\textbf {\bibinfo {volume} {6}},\ \bibinfo
  {pages} {1--6} (\bibinfo {year} {2016})}\BibitemShut {NoStop}%
\bibitem [{\citenamefont {Abrikosov}\ \emph {et~al.}(2012)\citenamefont
  {Abrikosov}, \citenamefont {Gorkov},\ and\ \citenamefont
  {Dzyaloshinski}}]{AGD}%
  \BibitemOpen
  \bibfield  {author} {\bibinfo {author} {\bibfnamefont {Aleksei~Alekseevich}\
  \bibnamefont {Abrikosov}}, \bibinfo {author} {\bibfnamefont {Lev~Petrovich}\
  \bibnamefont {Gorkov}}, \ and\ \bibinfo {author} {\bibfnamefont
  {Igor~Ekhielevich}\ \bibnamefont {Dzyaloshinski}},\ }\href@noop {} {\emph
  {\bibinfo {title} {Methods of quantum field theory in statistical physics}}}\
  (\bibinfo  {publisher} {Courier Corporation},\ \bibinfo {year}
  {2012})\BibitemShut {NoStop}%
\bibitem [{sm()}]{sm}%
  \BibitemOpen
  \href@noop {} {\bibinfo  {journal} {See Supplemental Material at [URL will be
  inserted by publisher] for details of the 2D and 3D models, their symmetries,
  and the Green's function methodology.}\ }\BibitemShut {NoStop}%
\bibitem [{\citenamefont {Young}\ and\ \citenamefont
  {Kane}(2015)}]{Young-Kane2015}%
  \BibitemOpen
\bibfield  {journal} {  }\bibfield  {author} {\bibinfo {author} {\bibfnamefont
  {Steve~M}\ \bibnamefont {Young}}\ and\ \bibinfo {author} {\bibfnamefont
  {Charles~L}\ \bibnamefont {Kane}},\ }\bibfield  {title} {\enquote {\bibinfo
  {title} {Dirac semimetals in two dimensions},}\ }\href@noop {} {\bibfield
  {journal} {\bibinfo  {journal} {Phys. Rev. Lett.}\ }\textbf {\bibinfo
  {volume} {115}},\ \bibinfo {pages} {126803} (\bibinfo {year}
  {2015})}\BibitemShut {NoStop}%
\bibitem [{\citenamefont {Hatsugai}\ and\ \citenamefont
  {Kohmoto}(1992)}]{HK1992}%
  \BibitemOpen
  \bibfield  {author} {\bibinfo {author} {\bibfnamefont {Yasuhiro}\
  \bibnamefont {Hatsugai}}\ and\ \bibinfo {author} {\bibfnamefont {Mahito}\
  \bibnamefont {Kohmoto}},\ }\bibfield  {title} {\enquote {\bibinfo {title}
  {Exactly solvable model of correlated lattice electrons in any dimensions},}\
  }\href@noop {} {\bibfield  {journal} {\bibinfo  {journal} {Journal of the
  Physical Society of Japan}\ }\textbf {\bibinfo {volume} {61}},\ \bibinfo
  {pages} {2056--2069} (\bibinfo {year} {1992})}\BibitemShut {NoStop}%
\bibitem [{\citenamefont {Phillips}\ \emph {et~al.}(2018)\citenamefont
  {Phillips}, \citenamefont {Setty},\ and\ \citenamefont {Zhang}}]{Setty2018}%
  \BibitemOpen
  \bibfield  {author} {\bibinfo {author} {\bibfnamefont {Philip~W}\
  \bibnamefont {Phillips}}, \bibinfo {author} {\bibfnamefont {Chandan}\
  \bibnamefont {Setty}}, \ and\ \bibinfo {author} {\bibfnamefont {Shuyi}\
  \bibnamefont {Zhang}},\ }\bibfield  {title} {\enquote {\bibinfo {title}
  {Absence of a charge diffusion pole at finite energies in an exactly solvable
  interacting flat-band model in d dimensions},}\ }\href@noop {} {\bibfield
  {journal} {\bibinfo  {journal} {Phys. Rev. B}\ }\textbf {\bibinfo {volume}
  {97}},\ \bibinfo {pages} {195102} (\bibinfo {year} {2018})}\BibitemShut
  {NoStop}%
\bibitem [{\citenamefont {Setty}(2020)}]{Setty2020}%
  \BibitemOpen
  \bibfield  {author} {\bibinfo {author} {\bibfnamefont {Chandan}\ \bibnamefont
  {Setty}},\ }\bibfield  {title} {\enquote {\bibinfo {title} {Pairing
  instability on a {L}uttinger surface: A non-{F}ermi liquid to superconductor
  transition and its {S}achdev-{Y}e-{K}itaev dual},}\ }\href@noop {} {\bibfield
   {journal} {\bibinfo  {journal} {Phys. Rev. B}\ }\textbf {\bibinfo {volume}
  {101}},\ \bibinfo {pages} {184506} (\bibinfo {year} {2020})}\BibitemShut
  {NoStop}%
\bibitem [{\citenamefont {Setty}(2021{\natexlab{a}})}]{Setty2021}%
  \BibitemOpen
  \bibfield  {author} {\bibinfo {author} {\bibfnamefont {Chandan}\ \bibnamefont
  {Setty}},\ }\bibfield  {title} {\enquote {\bibinfo {title} {Superconductivity
  from {L}uttinger surfaces: Emergent {S}achdev-{Y}e-{K}itaev physics with
  infinite-body interactions},}\ }\href@noop {} {\bibfield  {journal} {\bibinfo
   {journal} {Phys. Rev. B}\ }\textbf {\bibinfo {volume} {103}},\ \bibinfo
  {pages} {014501} (\bibinfo {year} {2021}{\natexlab{a}})}\BibitemShut
  {NoStop}%
\bibitem [{\citenamefont {Phillips}\ \emph {et~al.}(2020)\citenamefont
  {Phillips}, \citenamefont {Yeo},\ and\ \citenamefont {Huang}}]{Phillips2020}%
  \BibitemOpen
  \bibfield  {author} {\bibinfo {author} {\bibfnamefont {Philip~W}\
  \bibnamefont {Phillips}}, \bibinfo {author} {\bibfnamefont {Luke}\
  \bibnamefont {Yeo}}, \ and\ \bibinfo {author} {\bibfnamefont {Edwin~W}\
  \bibnamefont {Huang}},\ }\bibfield  {title} {\enquote {\bibinfo {title}
  {Exact theory for superconductivity in a doped {M}ott insulator},}\
  }\href@noop {} {\bibfield  {journal} {\bibinfo  {journal} {Nature Physics}\
  }\textbf {\bibinfo {volume} {16}},\ \bibinfo {pages} {1175--1180} (\bibinfo
  {year} {2020})}\BibitemShut {NoStop}%
\bibitem [{\citenamefont {Huang}\ \emph {et~al.}(2022)\citenamefont {Huang},
  \citenamefont {Nave},\ and\ \citenamefont {Phillips}}]{Phillips2021}%
  \BibitemOpen
  \bibfield  {author} {\bibinfo {author} {\bibfnamefont {Edwin~W}\ \bibnamefont
  {Huang}}, \bibinfo {author} {\bibfnamefont {Gabriele~La}\ \bibnamefont
  {Nave}}, \ and\ \bibinfo {author} {\bibfnamefont {Philip~W}\ \bibnamefont
  {Phillips}},\ }\bibfield  {title} {\enquote {\bibinfo {title} {Discrete
  symmetry breaking defines the {M}ott quartic fixed point},}\ }\href@noop {}
  {\bibfield  {journal} {\bibinfo  {journal} {Nat. Phys.}\ }\textbf {\bibinfo
  {volume} {18}},\ \bibinfo {pages} {511--516} (\bibinfo {year}
  {2022})}\BibitemShut {NoStop}%
\bibitem [{\citenamefont {Yang}(2021)}]{Yang2021}%
  \BibitemOpen
  \bibfield  {author} {\bibinfo {author} {\bibfnamefont {Kun}\ \bibnamefont
  {Yang}},\ }\bibfield  {title} {\enquote {\bibinfo {title} {Exactly solvable
  model of {F}ermi arcs and pseudogap},}\ }\href@noop {} {\bibfield  {journal}
  {\bibinfo  {journal} {Phys. Rev. B}\ }\textbf {\bibinfo {volume} {103}},\
  \bibinfo {pages} {024529} (\bibinfo {year} {2021})}\BibitemShut {NoStop}%
\bibitem [{\citenamefont {Zhu}\ \emph {et~al.}(2021)\citenamefont {Zhu},
  \citenamefont {Li}, \citenamefont {Han},\ and\ \citenamefont
  {Wang}}]{Wang2021}%
  \BibitemOpen
  \bibfield  {author} {\bibinfo {author} {\bibfnamefont {Huai-Shuang}\
  \bibnamefont {Zhu}}, \bibinfo {author} {\bibfnamefont {Zhidan}\ \bibnamefont
  {Li}}, \bibinfo {author} {\bibfnamefont {Qiang}\ \bibnamefont {Han}}, \ and\
  \bibinfo {author} {\bibfnamefont {ZD}~\bibnamefont {Wang}},\ }\bibfield
  {title} {\enquote {\bibinfo {title} {Topological s-wave superconductors
  driven by electron correlation},}\ }\href@noop {} {\bibfield  {journal}
  {\bibinfo  {journal} {Phys. Rev. B}\ }\textbf {\bibinfo {volume} {103}},\
  \bibinfo {pages} {024514} (\bibinfo {year} {2021})}\BibitemShut {NoStop}%
\bibitem [{\citenamefont {Setty}(2021{\natexlab{b}})}]{Setty2021-Kondo}%
  \BibitemOpen
  \bibfield  {author} {\bibinfo {author} {\bibfnamefont {Chandan}\ \bibnamefont
  {Setty}},\ }\bibfield  {title} {\enquote {\bibinfo {title} {Dilute magnetic
  moments in an exactly solvable interacting host},}\ }\href@noop {} {\bibfield
   {journal} {\bibinfo  {journal} {arXiv preprint arXiv:2105.15205}\ }
  (\bibinfo {year} {2021}{\natexlab{b}})}\BibitemShut {NoStop}%
\bibitem [{\citenamefont {Fabrizio}(2022)}]{Fabrizio2022}%
  \BibitemOpen
  \bibfield  {author} {\bibinfo {author} {\bibfnamefont {Michele}\ \bibnamefont
  {Fabrizio}},\ }\bibfield  {title} {\enquote {\bibinfo {title} {Emergent
  quasiparticles at {L}uttinger surfaces},}\ }\href@noop {} {\bibfield
  {journal} {\bibinfo  {journal} {Nature communications}\ }\textbf {\bibinfo
  {volume} {13}},\ \bibinfo {pages} {1--6} (\bibinfo {year}
  {2022})}\BibitemShut {NoStop}%
\bibitem [{\citenamefont {Raghu}\ \emph {et~al.}(2008)\citenamefont {Raghu},
  \citenamefont {Qi}, \citenamefont {Honerkamp},\ and\ \citenamefont
  {Zhang}}]{raghu2008topological}%
  \BibitemOpen
  \bibfield  {author} {\bibinfo {author} {\bibfnamefont {Srinivas}\
  \bibnamefont {Raghu}}, \bibinfo {author} {\bibfnamefont {Xiao-Liang}\
  \bibnamefont {Qi}}, \bibinfo {author} {\bibfnamefont {Carsten}\ \bibnamefont
  {Honerkamp}}, \ and\ \bibinfo {author} {\bibfnamefont {Shou-Cheng}\
  \bibnamefont {Zhang}},\ }\bibfield  {title} {\enquote {\bibinfo {title}
  {Topological {M}ott insulators},}\ }\href@noop {} {\bibfield  {journal}
  {\bibinfo  {journal} {Phys. Rev. Lett.}\ }\textbf {\bibinfo {volume} {100}},\
  \bibinfo {pages} {156401} (\bibinfo {year} {2008})}\BibitemShut {NoStop}%
\bibitem [{\citenamefont {Sheng}\ \emph {et~al.}(2011)\citenamefont {Sheng},
  \citenamefont {Gu}, \citenamefont {Sun},\ and\ \citenamefont
  {Sheng}}]{sheng2011fractional}%
  \BibitemOpen
  \bibfield  {author} {\bibinfo {author} {\bibfnamefont {DN}~\bibnamefont
  {Sheng}}, \bibinfo {author} {\bibfnamefont {Zheng-Cheng}\ \bibnamefont {Gu}},
  \bibinfo {author} {\bibfnamefont {Kai}\ \bibnamefont {Sun}}, \ and\ \bibinfo
  {author} {\bibfnamefont {L}~\bibnamefont {Sheng}},\ }\bibfield  {title}
  {\enquote {\bibinfo {title} {Fractional quantum hall effect in the absence of
  landau levels},}\ }\href@noop {} {\bibfield  {journal} {\bibinfo  {journal}
  {Nature communications}\ }\textbf {\bibinfo {volume} {2}},\ \bibinfo {pages}
  {389} (\bibinfo {year} {2011})}\BibitemShut {NoStop}%
\bibitem [{\citenamefont {Kang}\ and\ \citenamefont
  {Vafek}(2019)}]{kang2019strong}%
  \BibitemOpen
  \bibfield  {author} {\bibinfo {author} {\bibfnamefont {Jian}\ \bibnamefont
  {Kang}}\ and\ \bibinfo {author} {\bibfnamefont {Oskar}\ \bibnamefont
  {Vafek}},\ }\bibfield  {title} {\enquote {\bibinfo {title} {Strong coupling
  phases of partially filled twisted bilayer graphene narrow bands},}\
  }\href@noop {} {\bibfield  {journal} {\bibinfo  {journal} {Phys. Rev. Lett.}\
  }\textbf {\bibinfo {volume} {122}},\ \bibinfo {pages} {246401} (\bibinfo
  {year} {2019})}\BibitemShut {NoStop}%
\bibitem [{\citenamefont {Fu}\ \emph {et~al.}(2007)\citenamefont {Fu},
  \citenamefont {Kane},\ and\ \citenamefont {Mele}}]{fu2007PRL}%
  \BibitemOpen
  \bibfield  {author} {\bibinfo {author} {\bibfnamefont {Liang}\ \bibnamefont
  {Fu}}, \bibinfo {author} {\bibfnamefont {Charles~L}\ \bibnamefont {Kane}}, \
  and\ \bibinfo {author} {\bibfnamefont {Eugene~J}\ \bibnamefont {Mele}},\
  }\bibfield  {title} {\enquote {\bibinfo {title} {Topological insulators in
  three dimensions},}\ }\href@noop {} {\bibfield  {journal} {\bibinfo
  {journal} {Phys. Rev. Lett.}\ }\textbf {\bibinfo {volume} {98}},\ \bibinfo
  {pages} {106803} (\bibinfo {year} {2007})}\BibitemShut {NoStop}%
\bibitem [{\citenamefont {Fu}\ and\ \citenamefont {Kane}(2007)}]{fu2007PRB}%
  \BibitemOpen
  \bibfield  {author} {\bibinfo {author} {\bibfnamefont {Liang}\ \bibnamefont
  {Fu}}\ and\ \bibinfo {author} {\bibfnamefont {Charles~L}\ \bibnamefont
  {Kane}},\ }\bibfield  {title} {\enquote {\bibinfo {title} {Topological
  insulators with inversion symmetry},}\ }\href@noop {} {\bibfield  {journal}
  {\bibinfo  {journal} {Phys. Rev. B}\ }\textbf {\bibinfo {volume} {76}},\
  \bibinfo {pages} {045302} (\bibinfo {year} {2007})}\BibitemShut {NoStop}%
\bibitem [{\citenamefont {Setty}\ \emph
  {et~al.}(2023{\natexlab{a}})\citenamefont {Setty}, \citenamefont {Xie},
  \citenamefont {Sur}, \citenamefont {Chen}, \citenamefont {Paschen},
  \citenamefont {Vergniory}, \citenamefont {Cano},\ and\ \citenamefont
  {Si}}]{setty2023topological}%
  \BibitemOpen
  \bibfield  {author} {\bibinfo {author} {\bibfnamefont {Chandan}\ \bibnamefont
  {Setty}}, \bibinfo {author} {\bibfnamefont {Fang}\ \bibnamefont {Xie}},
  \bibinfo {author} {\bibfnamefont {Shouvik}\ \bibnamefont {Sur}}, \bibinfo
  {author} {\bibfnamefont {Lei}\ \bibnamefont {Chen}}, \bibinfo {author}
  {\bibfnamefont {Silke}\ \bibnamefont {Paschen}}, \bibinfo {author}
  {\bibfnamefont {Maia~G}\ \bibnamefont {Vergniory}}, \bibinfo {author}
  {\bibfnamefont {Jennifer}\ \bibnamefont {Cano}}, \ and\ \bibinfo {author}
  {\bibfnamefont {Qimiao}\ \bibnamefont {Si}},\ }\bibfield  {title} {\enquote
  {\bibinfo {title} {Topological diagnosis of strongly correlated electron
  systems},}\ }\href@noop {} {\bibfield  {journal} {\bibinfo  {journal} {arXiv
  preprint arXiv:2311.12031}\ } (\bibinfo {year}
  {2023}{\natexlab{a}})}\BibitemShut {NoStop}%
\bibitem [{\citenamefont {Setty}\ \emph
  {et~al.}(2023{\natexlab{b}})\citenamefont {Setty}, \citenamefont {Xie},
  \citenamefont {Sur}, \citenamefont {Chen}, \citenamefont {Vergniory},\ and\
  \citenamefont {Si}}]{setty2023electronic}%
  \BibitemOpen
  \bibfield  {author} {\bibinfo {author} {\bibfnamefont {Chandan}\ \bibnamefont
  {Setty}}, \bibinfo {author} {\bibfnamefont {Fang}\ \bibnamefont {Xie}},
  \bibinfo {author} {\bibfnamefont {Shouvik}\ \bibnamefont {Sur}}, \bibinfo
  {author} {\bibfnamefont {Lei}\ \bibnamefont {Chen}}, \bibinfo {author}
  {\bibfnamefont {Maia~G}\ \bibnamefont {Vergniory}}, \ and\ \bibinfo {author}
  {\bibfnamefont {Qimiao}\ \bibnamefont {Si}},\ }\bibfield  {title} {\enquote
  {\bibinfo {title} {Electronic properties, correlated topology and {G}reen's
  function zeros},}\ }\href@noop {} {\bibfield  {journal} {\bibinfo  {journal}
  {arXiv preprint arXiv:2309.14340}\ } (\bibinfo {year}
  {2023}{\natexlab{b}})}\BibitemShut {NoStop}%
\bibitem [{\citenamefont {Wieder}\ \emph {et~al.}(2016)\citenamefont {Wieder},
  \citenamefont {Kim}, \citenamefont {Rappe},\ and\ \citenamefont
  {Kane}}]{Wieder2016}%
  \BibitemOpen
  \bibfield  {author} {\bibinfo {author} {\bibfnamefont {Benjamin~J.}\
  \bibnamefont {Wieder}}, \bibinfo {author} {\bibfnamefont {Youngkuk}\
  \bibnamefont {Kim}}, \bibinfo {author} {\bibfnamefont {A.~M.}\ \bibnamefont
  {Rappe}}, \ and\ \bibinfo {author} {\bibfnamefont {C.~L.}\ \bibnamefont
  {Kane}},\ }\bibfield  {title} {\enquote {\bibinfo {title} {Double dirac
  semimetals in three dimensions},}\ }\href@noop {} {\bibfield  {journal} {\bibinfo  {journal}
  {Phys. Rev. Lett.}\ }\textbf {\bibinfo {volume} {116}},\ \bibinfo {pages}
  {186402} (\bibinfo {year} {2016})}\BibitemShut {NoStop}%
\bibitem [{\citenamefont {Bradlyn}\ \emph {et~al.}(2016)\citenamefont
  {Bradlyn}, \citenamefont {Cano}, \citenamefont {Wang}, \citenamefont
  {Vergniory}, \citenamefont {Felser}, \citenamefont {Cava},\ and\
  \citenamefont {Bernevig}}]{Bradlyn2016}%
  \BibitemOpen
  \bibfield  {author} {\bibinfo {author} {\bibfnamefont {Barry}\ \bibnamefont
  {Bradlyn}}, \bibinfo {author} {\bibfnamefont {Jennifer}\ \bibnamefont
  {Cano}}, \bibinfo {author} {\bibfnamefont {Zhijun}\ \bibnamefont {Wang}},
  \bibinfo {author} {\bibfnamefont {M.~G.}\ \bibnamefont {Vergniory}}, \bibinfo
  {author} {\bibfnamefont {C.}~\bibnamefont {Felser}}, \bibinfo {author}
  {\bibfnamefont {R.~J.}\ \bibnamefont {Cava}}, \ and\ \bibinfo {author}
  {\bibfnamefont {B.~Andrei}\ \bibnamefont {Bernevig}},\ }\bibfield  {title}
  {\enquote {\bibinfo {title} {Beyond {D}irac and {W}eyl {F}ermions: Unconventional
  quasiparticles in conventional crystals},}\ }\href@noop {} {\bibfield  {journal} {\bibinfo  {journal}
  {Science}\ }\textbf {\bibinfo {volume} {353}},\ \bibinfo {pages} {aaf5037}
  (\bibinfo {year} {2016})}\BibitemShut {NoStop}%
\bibitem [{\citenamefont {Di~Sante}\ \emph {et~al.}(2017)\citenamefont
  {Di~Sante}, \citenamefont {Hausoel}, \citenamefont {Barone}, \citenamefont
  {Tomczak}, \citenamefont {Sangiovanni},\ and\ \citenamefont
  {Thomale}}]{DiSante2017}%
  \BibitemOpen
  \bibfield  {author} {\bibinfo {author} {\bibfnamefont {Domenico}\
  \bibnamefont {Di~Sante}}, \bibinfo {author} {\bibfnamefont {Andreas}\
  \bibnamefont {Hausoel}}, \bibinfo {author} {\bibfnamefont {Paolo}\
  \bibnamefont {Barone}}, \bibinfo {author} {\bibfnamefont {Jan~M.}\
  \bibnamefont {Tomczak}}, \bibinfo {author} {\bibfnamefont {Giorgio}\
  \bibnamefont {Sangiovanni}}, \ and\ \bibinfo {author} {\bibfnamefont {Ronny}\
  \bibnamefont {Thomale}},\ }\bibfield  {title} {\enquote {\bibinfo {title}
  {Realizing double dirac particles in the presence of electronic
  interactions},}\ }\href@noop {} {\bibfield
  {journal} {\bibinfo  {journal} {Phys. Rev. B}\ }\textbf {\bibinfo {volume}
  {96}},\ \bibinfo {pages} {121106} (\bibinfo {year} {2017})}\BibitemShut
  {NoStop}%
\bibitem [{\citenamefont {Goldoni}\ \emph {et~al.}(1994)\citenamefont
  {Goldoni}, \citenamefont {del Pennino}, \citenamefont {Parmigiani},
  \citenamefont {Sangaletti},\ and\ \citenamefont
  {Revcolevschi}}]{goldoni1994electronic}%
  \BibitemOpen
  \bibfield  {author} {\bibinfo {author} {\bibfnamefont {A}~\bibnamefont
  {Goldoni}}, \bibinfo {author} {\bibfnamefont {Umberto}\ \bibnamefont {del
  Pennino}}, \bibinfo {author} {\bibfnamefont {F}~\bibnamefont {Parmigiani}},
  \bibinfo {author} {\bibfnamefont {L}~\bibnamefont {Sangaletti}}, \ and\
  \bibinfo {author} {\bibfnamefont {A}~\bibnamefont {Revcolevschi}},\
  }\bibfield  {title} {\enquote {\bibinfo {title} {Electronic structure of
  {Bi$_2$CuO$_4$}},}\ }\href@noop {} {\bibfield  {journal} {\bibinfo  {journal}
  {Phys. Rev. B}\ }\textbf {\bibinfo {volume} {50}},\ \bibinfo {pages} {10435}
  (\bibinfo {year} {1994})}\BibitemShut {NoStop}%
\bibitem [{\citenamefont {Garcia-Munoz}\ \emph {et~al.}(1990)\citenamefont
  {Garcia-Munoz}, \citenamefont {Rodriguez-Carvajal}, \citenamefont {Sapina},
  \citenamefont {Sanchis}, \citenamefont {Ibanez},\ and\ \citenamefont
  {Beltran-Porter}}]{garcia1990crystal}%
  \BibitemOpen
  \bibfield  {author} {\bibinfo {author} {\bibfnamefont {JL}~\bibnamefont
  {Garcia-Munoz}}, \bibinfo {author} {\bibfnamefont {J}~\bibnamefont
  {Rodriguez-Carvajal}}, \bibinfo {author} {\bibfnamefont {F}~\bibnamefont
  {Sapina}}, \bibinfo {author} {\bibfnamefont {MJ}~\bibnamefont {Sanchis}},
  \bibinfo {author} {\bibfnamefont {R}~\bibnamefont {Ibanez}}, \ and\ \bibinfo
  {author} {\bibfnamefont {D}~\bibnamefont {Beltran-Porter}},\ }\bibfield
  {title} {\enquote {\bibinfo {title} {Crystal and magnetic structures of
  {Bi$_2$CuO$_4$}},}\ }\href@noop {} {\bibfield  {journal} {\bibinfo  {journal}
  {Journal of Physics: Condensed Matter}\ }\textbf {\bibinfo {volume} {2}},\
  \bibinfo {pages} {2205} (\bibinfo {year} {1990})}\BibitemShut {NoStop}%
\bibitem [{\citenamefont {Chen}\ \emph
  {et~al.}(2022{\natexlab{b}})\citenamefont {Chen}, \citenamefont {Xie},
  \citenamefont {Sur}, \citenamefont {Hu}, \citenamefont {Paschen},
  \citenamefont {Cano},\ and\ \citenamefont {Si}}]{Chen-emergent2022}%
  \BibitemOpen
  \bibfield  {author} {\bibinfo {author} {\bibfnamefont {Lei}\ \bibnamefont
  {Chen}}, \bibinfo {author} {\bibfnamefont {Fang}\ \bibnamefont {Xie}},
  \bibinfo {author} {\bibfnamefont {Shouvik}\ \bibnamefont {Sur}}, \bibinfo
  {author} {\bibfnamefont {Haoyu}\ \bibnamefont {Hu}}, \bibinfo {author}
  {\bibfnamefont {Silke}\ \bibnamefont {Paschen}}, \bibinfo {author}
  {\bibfnamefont {Jennifer}\ \bibnamefont {Cano}}, \ and\ \bibinfo {author}
  {\bibfnamefont {Qimiao}\ \bibnamefont {Si}},\ }\bibfield  {title} {\enquote
  {\bibinfo {title} {Emergent flat band and topological {K}ondo semimetal
  driven by orbital-selective correlations},}\ }\href@noop {} {\bibfield
  {journal} {\bibinfo  {journal} {arXiv preprint arXiv:2212.08017}\ } (\bibinfo
  {year} {2022}{\natexlab{b}})},\ \bibinfo {note} {{N}at. {C}ommun., in press
  (2024)}\BibitemShut {NoStop}%
\bibitem [{\citenamefont {Hu}\ and\ \citenamefont {Si}(2023)}]{Hu-FB2022}%
  \BibitemOpen
  \bibfield  {author} {\bibinfo {author} {\bibfnamefont {Haoyu}\ \bibnamefont
  {Hu}}\ and\ \bibinfo {author} {\bibfnamefont {Qimiao}\ \bibnamefont {Si}},\
  }\bibfield  {title} {\enquote {\bibinfo {title} {Coupled topological flat and
  wide bands: Quasiparticle formation and destruction},}\ }\href@noop {} {\bibfield  {journal} {\bibinfo  {journal} {Sci.
  Adv.}\ }\textbf {\bibinfo {volume} {9}},\ \bibinfo {pages} {eadg0028}
  (\bibinfo {year} {2023})}\BibitemShut {NoStop}%
\bibitem [{\citenamefont {Chen}\ \emph {et~al.}(2023)\citenamefont {Chen},
  \citenamefont {Xie}, \citenamefont {Sur}, \citenamefont {Hu}, \citenamefont
  {Paschen}, \citenamefont {Cano},\ and\ \citenamefont {Si}}]{Che23.1x}%
  \BibitemOpen
  \bibfield  {author} {\bibinfo {author} {\bibfnamefont {Lei}\ \bibnamefont
  {Chen}}, \bibinfo {author} {\bibfnamefont {Fang}\ \bibnamefont {Xie}},
  \bibinfo {author} {\bibfnamefont {Shouvik}\ \bibnamefont {Sur}}, \bibinfo
  {author} {\bibfnamefont {Haoyu}\ \bibnamefont {Hu}}, \bibinfo {author}
  {\bibfnamefont {Silke}\ \bibnamefont {Paschen}}, \bibinfo {author}
  {\bibfnamefont {Jennifer}\ \bibnamefont {Cano}}, \ and\ \bibinfo {author}
  {\bibfnamefont {Qimiao}\ \bibnamefont {Si}},\ }\bibfield  {title} {\enquote
  {\bibinfo {title} {Metallic quantum criticality enabled by flat bands in a
  kagome lattice},}\ }\href@noop {} {\bibfield  {journal} {\bibinfo  {journal}
  {arXiv preprint arXiv:2307.09431}\ } (\bibinfo {year} {2023})}\BibitemShut
  {NoStop}%
\bibitem [{\citenamefont {Wagner}\ \emph {et~al.}(2023)\citenamefont {Wagner},
  \citenamefont {Crippa}, \citenamefont {Amaricci}, \citenamefont {Hansmann},
  \citenamefont {Klett}, \citenamefont {K{\"o}nig}, \citenamefont
  {Sch{\"a}fer}, \citenamefont {Sante}, \citenamefont {Cano}, \citenamefont
  {Millis}, \citenamefont {Georges},\ and\ \citenamefont
  {Sangiovanni}}]{wagner2023mott}%
  \BibitemOpen
  \bibfield  {author} {\bibinfo {author} {\bibfnamefont {Niklas}\ \bibnamefont
  {Wagner}}, \bibinfo {author} {\bibfnamefont {Lorenzo}\ \bibnamefont
  {Crippa}}, \bibinfo {author} {\bibfnamefont {Adriano}\ \bibnamefont
  {Amaricci}}, \bibinfo {author} {\bibfnamefont {Philipp}\ \bibnamefont
  {Hansmann}}, \bibinfo {author} {\bibfnamefont {Marcel}\ \bibnamefont
  {Klett}}, \bibinfo {author} {\bibfnamefont {EJ}~\bibnamefont {K{\"o}nig}},
  \bibinfo {author} {\bibfnamefont {Thomas}\ \bibnamefont {Sch{\"a}fer}},
  \bibinfo {author} {\bibfnamefont {D~Di}\ \bibnamefont {Sante}}, \bibinfo
  {author} {\bibfnamefont {Jennifer}\ \bibnamefont {Cano}}, \bibinfo {author}
  {\bibfnamefont {AJ}~\bibnamefont {Millis}}, \bibinfo {author} {\bibfnamefont
  {Antoine}\ \bibnamefont {Georges}}, \ and\ \bibinfo {author} {\bibfnamefont
  {Giorgio}\ \bibnamefont {Sangiovanni}},\ }\bibfield  {title} {\enquote
  {\bibinfo {title} {{M}ott insulators with boundary zeros},}\ }\href@noop {}
  {\bibfield  {journal} {\bibinfo  {journal} {Nature Communications}\ }\textbf
  {\bibinfo {volume} {14}},\ \bibinfo {pages} {7531} (\bibinfo {year}
  {2023})}\BibitemShut {NoStop}%
\end{thebibliography}%

 \end{document}